\documentclass[twocolumn,trackchanges]{aastex62}
\usepackage{booktabs}
\usepackage{color}
\usepackage{soul}
\usepackage{ulem}
\usepackage{bm}
\usepackage{amsmath}

\hypersetup{linkcolor=red,citecolor=blue,filecolor=cyan,urlcolor=magenta}
\submitjournal{ApJS}

\shortauthors{Song et al.}
\begin{document}

\title{Large-scale multiconfiguration Dirac-Hartree-Fock calculations for astrophysics:
n=4 levels in  P-like ions from  Mn~XI to Ni~XIV}

\author{C.~X. Song}
\affil{Hebei Key Lab of Optic-electronic Information and Materials, The College of Physics Science and Technology, Hebei University, Baoding 071002, China; wang$_{-}$kai10@fudan.edu.cn}

\author{K. Wang}
\affil{Hebei Key Lab of Optic-electronic Information and Materials, The College of Physics Science and Technology, Hebei University, Baoding 071002, China; wang$_{-}$kai10@fudan.edu.cn}
\affil{Shanghai EBIT Lab, Key Laboratory of Nuclear Physics and Ion-beam Application, Institute of Modern Physics, Department of Nuclear Science and Technology, Fudan University, Shanghai 200433, China;}

\author{G. Del Zanna}
\affil{DAMTP, Centre for Mathematical Sciences, University of Cambridge, Wilberforce Road, Cambridge CB3 0WA, UK}

\author{P. J\"onsson}
\affil{Group for Materials Science and Applied Mathematics, Malm\"o University, SE-20506, Malm\"o, Sweden; \\}

\author{R. Si}
\affil{Spectroscopy, Quantum Chemistry and Atmospheric Remote Sensing, CP160/09, Universit\'{e} libre de Bruxelles,  1050 Brussels, Belgium;}

\author{M. Godefroid}
\affil{Spectroscopy, Quantum Chemistry and Atmospheric Remote Sensing, CP160/09, Universit\'{e} libre de Bruxelles,  1050 Brussels, Belgium;}

\author{G. Gaigalas}
\affil{Institute of Theoretical Physics and Astronomy, Vilnius University, Saul\.etekio av. 3, LT-10222, Vilnius, Lithuania;}

\author{L. Rad{\v{z}}i{\={u}}t{\.{e}}} 
\affil{Institute of Theoretical Physics and Astronomy, Vilnius University, Saul\.etekio av. 3, LT-10222, Vilnius, Lithuania;}

\author{P. Rynkun}
\affil{Institute of Theoretical Physics and Astronomy, Vilnius University, Saul\.etekio av. 3, LT-10222, Vilnius, Lithuania;}

\author{X. H. Zhao}
\affil{Hebei Key Lab of Optic-electronic Information and Materials, The College of Physics Science and Technology, Hebei University, Baoding 071002, China; wang$_{-}$kai10@fudan.edu.cn}

\author{J. Yan}
\affil{Institute of Applied Physics and Computational Mathematics, Beijing 100088, China;}
\author{C.~Y. Chen}
\affil{Shanghai EBIT Lab, Key Laboratory of Nuclear Physics and Ion-beam Application, Institute of Modern Physics, Department of Nuclear Science and Technology, Fudan University, Shanghai 200433, China;}

\nocollaboration



\begin{abstract}
Using the multiconfiguration Dirac-Hartree-Fock and the relativistic configuration interaction methods,
  a consistent set of transition energies and radiative transition data for the lowest 546 (623, 701, 745)
  states of the $3p^4 3d$, $3s 3p^2 3d^2$, $3s 3p^3 4p$,  $3s 3p^4$, $3s^2 3d^3$, $3s^2 3p^2 3d$, $3s^2 3p^2 4d$, $3s^2 3p^2 4s$, $3p^3 3d^2$, $3p^5$, $3s 3p 3d^3$,  $3s 3p^3 3d$, $3s 3p^3 4s$, $3s^2 3p 3d^2$, %$3s^2 3p^2 4f$, 
$3s^2 3p^2 4p$, $3s^2 3p^3$ configurations in  Mn~XI (Fe~XII, Co~XIII, Ni~XIV) is provided. 
  The comparison between calculated excitation energies for the $n=4$ states and available experimental values for \ion{Fe}{12}
  indicate that  the calculations are highly accurate, with uncertainties of only a few hundred cm$^{-1}$.
  Lines from these states are prominent in the soft X-rays.
  With the present calculations, several recent new identifications are confirmed. 
  Other identifications  involving $3p^2 4d$ levels in Fe~XII that were found questionable are discussed and a few new assignments are recommended.
  As some  $n=4$ states of the other ions also show large discrepancies between experimental and
  calculated energies, we reassess their identification.
  The present study  provides highly accurate atomic data for the $n=4$ states
  of  P-like ions of astrophysical interest, for which experimental data are scarce.
\end{abstract}




\section{Introduction} \label{sec:intro}

  In the ultraviolet (UV) and extreme ultraviolet (EUV) spectral regions, P-like ions of the iron group
  elements are used for plasma diagnostics,
  especially to measure electron densities~\citep[see, e.g. the review by][]{Zanna.2018.V15.p}. 
\mbox{Fe XII} is especially important for the solar corona. 
The most prominent lines  in the UV/EUV region were observed by  SOHO and Skylab,
and were recently measured by the Hinode satellite,
~\citep{Young.2009.V495.p587,DelZanna.2012.V537.p38}. 
In our recent  work~\citep{Wang.2018.V235.p27}, using the multiconfiguration Dirac--Hartree--Fock (MCDHF)
and the relativistic configuration interaction (RCI) methods~\citep{FroeseFischer.2016.V49.p182004,Grant.2007.V.p},
excitation energies for the lowest 143  states of the $n=3$  ($3s^23p^3$, $3s3p^4$, $3s^23p^23d$, $3s3p^33d$, $3p^5$, $3s^23p3d^2$)
configurations  from \ion{Cr}{10} to \ion{Zn}{16} were provided. 
Using these data, we reassessed the previous identifications of the important $3s^2 3p^2 3d$ levels in \ion{Fe}{12},
confirming most of the previous suggestions by  \cite{delzanna_mason:05_fe_12}.
On the basis of our calculated energies, the atomic data from \cite{Zanna.2012.V541.p90},
and Hinode EIS spectra, new identifications of a few  $3s 3p^3 3d$ levels were also suggested,
which have been included in the CHIANTI atomic database version 9
 \citep{Dere.2019.V241.p22,Dere.1997.V125.p149}.

The $n=4 \to n=3$ transitions in highly-ionized Fe ions, including \mbox{Fe XII},
  are the dominant lines in the soft X-rays (50~\AA -- 150~\AA). 
  These transitions have recently been reviewed by~\citet{DelZanna.2012.V546.p97},
  where  a series of large-scale scattering calculations (the first of the kind) for all
  the ions, provided by Del Zanna et al. in a series of papers, 
  were used to assess the identifications in this spectral region. 
  Most of the previous identifications of these $n=4 \to n=3$ transitions were due to~\citet{Fawcett.1972.V5.p1255},
  although some of the original ones were due to Edlen in his seminal work in the 1930s'
  (see, e.g. the review in~\citet{Zanna.2018.V15.p}). 
  \citet{DelZanna.2012.V546.p97} pointed out that several of the strongest transitions were never identified, and proposed their identification, on the basis  of solar and laboratory spectra (the same used by~\citet{Fawcett.1972.V5.p1255}). A few problems with  Fawcett's identifications were also noted, but the accuracy of the theoretical  wavelengths did not allow firm identifications. 
  
A significant fraction (about a third) of the spectral lines in the soft X-rays still awaits identification.
One purpose of  our work is to provide excitation energies and wavelengths involving the $n=4$ states for P-like ions from \ion{Mn}{11} to \ion{Ni}{14} with spectroscopic accuracy, to aid the identification process.
Using the MCDHF and RCI method (in the following referred to as MCDHF), excitation energies, wavelengths, lifetimes, and radiative transition data including oscillator strengths, line strengths, and transition rates, are provided for the main $n=3,4$ levels of the $3p^4 3d$, $3s 3p^2 3d^2$, $3s 3p^3 4p$,  $3s 3p^4$, $3s^2 3d^3$, $3s^2 3p^2 3d$, $3s^2 3p^2 4d$, $3s^2 3p^2 4s$, $3p^3 3d^2$, $3p^5$, $3s 3p 3d^3$,  $3s 3p^3 3d$, $3s 3p^3 4s$, $3s^2 3p 3d^2$, %$3s^2 3p^2 4f$, 
$3s^2 3p^2 4p$, $3s^2 3p^3$ configurations.
In Section~\ref{Sec:evaluation}, using accurate wavelengths for the $n=4 \to n=3$ transitions of \ion{Fe}{12} to \ion{Ni}{14} we will review their identification and suggest some new lines. This provides a stringent accuracy assessment of our calculations.

The other purpose of  our work is to provide a consistent accurate set of radiative transition data for P-like ions from \ion{Mn}{11} to \ion{Ni}{14} for spectral line modeling. This work extends and complements our long-term theoretical efforts~\citep{Wang.2014.V215.p26,Wang.2015.V218.p16,Wang.2016.V223.p3,Wang.2016.V226.p14,Wang.2017.V119.p189301,Wang.2017.V194.p108,Wang.2017.V187.p375,Wang.2017.V229.p37,Wang.2018.V235.p27,Wang.2018.V239.p30,Wang.2018.V234.p40,Wang.2018.V208.p134,Wang.2019.V236.p106586,Wang.2020.V246.p1,Chen.2017.V113.p258,Chen.2018.V206.p213,Guo.2015.V48.p144020,Guo.2016.V93.p12513,Si.2016.V227.p16,Si.2017.V189.p249,Si.2018.V239.p3,Zhao.2018.V119.p314} to provide atomic data for L- and M-shells systems with high accuracy.
For a review see \cite{Jonsson.2017.V5.p16}.

\section{Theory and Calculations}
The MCDHF method in the GRASP2K code\\~\citep{Jonsson.2013.V184.p2197,Jonsson.2007.V177.p597} is reviewed  by~\citet{FroeseFischer.2016.V49.p182004}. This method is also described in our recent papers~\citep{Wang.2018.V235.p27,Wang.2018.V239.p30}. For this reason, in the sections below, only the computational procedures are described.

%\subsection{MCDHF}\label{Sec:MCDHF}
In our MCDHF calculations, the multireference (MR) sets for even and odd parities include
%\begin{itemize}
%\item [even %configurations
%:]

even configurations:
$3p^4 3d$, $3s 3p^2 3d^2$, $3s 3p^3 4p$,  $3s 3p^4$, $3s^2 3d^3$, $3s^2 3p^2 3d$, $3s^2 3p^2 4d$, $3s^2 3p^2 4s$;
%\item [odd  %configurations
%:]

odd  configurations:
$3p^3 3d^2$, $3p^5$, $3s 3p 3d^3$,  $3s 3p^3 3d$, $3s 3p^3 4s$, $3s^2 3p 3d^2$, $3s^2 3p^2 4f$, $3s^2 3p^2 4p$, $3s^2 3p^3$.
%\end{itemize}

By allowing single and double  substitutions from the $n=3,4$ electrons of the MR sets to orbitals with  $n\leq8, l\leq6$, and allowing single excitations of the $n=2$ electrons to orbitals with $n\leq6, l\leq4$,  configuration state function (CSF) expansions are generated. 
The $1s$ shell is defined as inactive closed shell in all CSFs of the expansions. 

For both energy separations and  transition probabilities, the neglected correlations from $n=1,2$ are comparatively unimportant~\citep{Wang.2018.V235.p27}. 
In the following RCI calculation, the transverse electron interaction in the low-frequency limit and the  leading quantum electrodynamic (QED) effects (vacuum polarization and self-energy) corrections are included. 
In the final  CSF expansions for the different $J$ symmetries,  the number of CSFs  is, respectively, about 5.9 millions for even parity and 8.1 millions for odd parity. 

By  using the $jj$-$LSJ$ transformation approach\\~\citep{Gaigalas.2017.V5.p6, Gaigalas.2004.V157.p239}, the  $jj$-coupled CSFs are transformed into $LSJ$-coupled CSFs, from which the $LSJ$ labels used by experimentalists are obtained.

%\subsection{MBPT}
%In the  MBPT method, the Hilbert space of the full Hamiltonian is divided into two parts, i.e., a model space $M$ and a orthogonal space $N$.  We included all the CSFs of the MR set in the above MCDHF calculation in the $M$ space. All the possible CSFs generated by permitting single  and double  substitutions from the electrons of the MR sets are included in the $N$ space.  With the maximum $l$ value of 20, for single and double  excitations the maximum $n$ values are, respectively, 125 and 65.  The configuration interaction effects in the model space $M$ are considered in the self-consistent field calculations non-perturbatively.  The effects between the two spaces $M$ and $N$ are considered by using the second-order perturbation method. 

% GDZ: I dod not understand what this plot conveys. It would be more
% instructive perhaps to plot differneces in A-values as a function of the 
% line strength 

%\begin{figure*}[ht!]
%	\plotone{../figure/fig.lf.fe.eps}
%		\plotone{fig.lf.fe.eps}
%	\caption{The differences ($\tau_{\rm other}/\tau_{\rm MCDHF}^l-1$) in~\% of the present  results $\tau_{\rm MCDHF}^v$ in the velocity form and the results $\tau_{\rm AS}^l$ in the length form from the {\sc autostructure}  calculations by~\citet{Zanna.2012.V541.p90}, relative to the present values $\tau_{\rm MCDHF}^l$  in length form.\label{fig.lf.ne}}
%\end{figure*}

\section{EVALUATION OF DATA}~\label{Sec:evaluation}

\subsection{Energy Levels}~\label{Sec:en}
In Table~\ref{tab.lev.fe} excitation energies for the lowest 623  levels of the $3p^4 3d$, $3s 3p^2 3d^2$, $3s 3p^3 4p$,  $3s 3p^4$, $3s^2 3d^3$, $3s^2 3p^2 3d$, $3s^2 3p^2 4d$, $3s^2 3p^2 4s$, $3p^3 3d^2$, $3p^5$, $3s 3p 3d^3$,  $3s 3p^3 3d$, $3s 3p^3 4s$, $3s^2 3p 3d^2$, %$3s^2 3p^2 4f$, 
$3s^2 3p^2 4p$, $3s^2 3p^3$ configurations in \mbox{Fe~XII} from the present MCDHF calculations (hereafter referred to as MCDHF1) are displayed. All these states are below the first $3s^2 3p^2 4f$ level.
For comparison, experimental excitation energies compiled in the Atomic Spectra Database (ASD) of the National Institute of Standards and Technology (NIST) ~\citep{Kramida.2018.V.p}, experimental (\mbox{CHIANTI1}) and calculated (CHIANTI2) values from the CHIANTI version~9~\citep{Dere.2019.V241.p22,Dere.1997.V125.p149} are also included along with excitation energies from our recently MCDHF calculations (MCDHF2)~\citep{Wang.2018.V235.p27}.  
The previous calculations, such as~\citet{Tayal.2011.V97.p481,Storey.2005.V433.p717,Vilkas.2004.V37.p4763}, were focused on the $n=3$ levels, and their comparison with our recently MCDHF results for the $n=3$ levels was shown in our previous work~\citep{Wang.2018.V235.p27}. 
Therefore, their results~\citep{Tayal.2011.V97.p481,Storey.2005.V433.p717,Vilkas.2004.V37.p4763} are  not included in Table \ref{tab.lev.fe}.

The differences between the MCDHF1 and MCDHF2 results for the lowest 143 states are listed in Table~\ref{tab.lev.fe}, and are generally within a few hundreds cm$^{-1}$. 
The average absolute difference with the standard deviation~\citep{Wang.2017.V229.p37}  between the two data sets is 1 cm$^{-1}$ $\pm$ 313 cm$^{-1}$. 
When the present MCDHF1 excitation energies for the $n=3$ levels are compared with the experimental values from the NIST and CHIANTI database, the agreement is also very good. The differences are generally within a few hundred cm$^{-1}$. There is a misprint for experimental excitation energies of the two levels ($3s^{2}\,3p^{2}(^{1}_{2}D)~^{1}D\,3d~^{2}P_{3/2}$ with the key ($\#37$) and $\#38 / 3s^{2}\,3p^{2}(^{1}_{2}D)~^{1}D\,3d~^{2}S_{1/2}$) in the CHIANTI version 9.  The experimental values should be 577 680 cm$^{-1}$ for the $\#37 / 3s^{2}\,3p^{2}(^{1}_{2}D)~^{1}D\,3d~^{2}P_{3/2}$ level, and 579 630 cm$^{-1}$ for the $\#38 / 3s^{2}\,3p^{2}(^{1}_{2}D)~^{1}D\,3d~^{2}S_{1/2}$ level, respectively~\citep{DelZanna.2005.V433.p731}. 
After correcting this misprint,  the average absolute differences with our MCDHF1 energy values are $-164 \pm 313$ cm$^{-1}$ for NIST  and $-113 \pm 313$ cm$^{-1}$ for \mbox{CHIANTI1}, respectively,  where the standard deviations are indicated after the values. 

Looking at higher-lying levels (above $\#143$) in \ion{Fe}{12}, the present MCDHF1 calculations, as well as the \mbox{CHIANTI2} theoretical values reported by the {\sc autostructure}  calculations by \citet{DelZanna.2012.V543.p139} provide a complete data set. 
The {\sc autostructure} calculations were carried out to provide target states wave functions for the scattering calculations. They are therefore not very accurate, providing indeed excitation energies that depart substantially from the present MCDHF1 values. 
For almost all higher-lying states of \ion{Fe}{12}, experimental values (NIST and \mbox{CHIANTI1}) are scarce. 
Only some levels of the $3s^2 3p^2 4s$,  $3s^2 3p^2 4p$, $3s^2 3p^2 4d$, and $3s 3p^3 4s$  were identified,
  based on the solar and laboratory spectra of \citet{Behring.1972.V175.p493,Fawcett.1972.V5.p1255}.
The differences between our MCDHF1 excitation energies and the experimental values (NIST and \mbox{CHIANTI1})
are generally within a few hundreds of  cm$^{-1}$. However, several discrepancies have been noted, as described below.

\begin{figure*}[ht!]
%		\plotone{../figure/fig.lev.nistlargedifferences.eps}
	\plotone{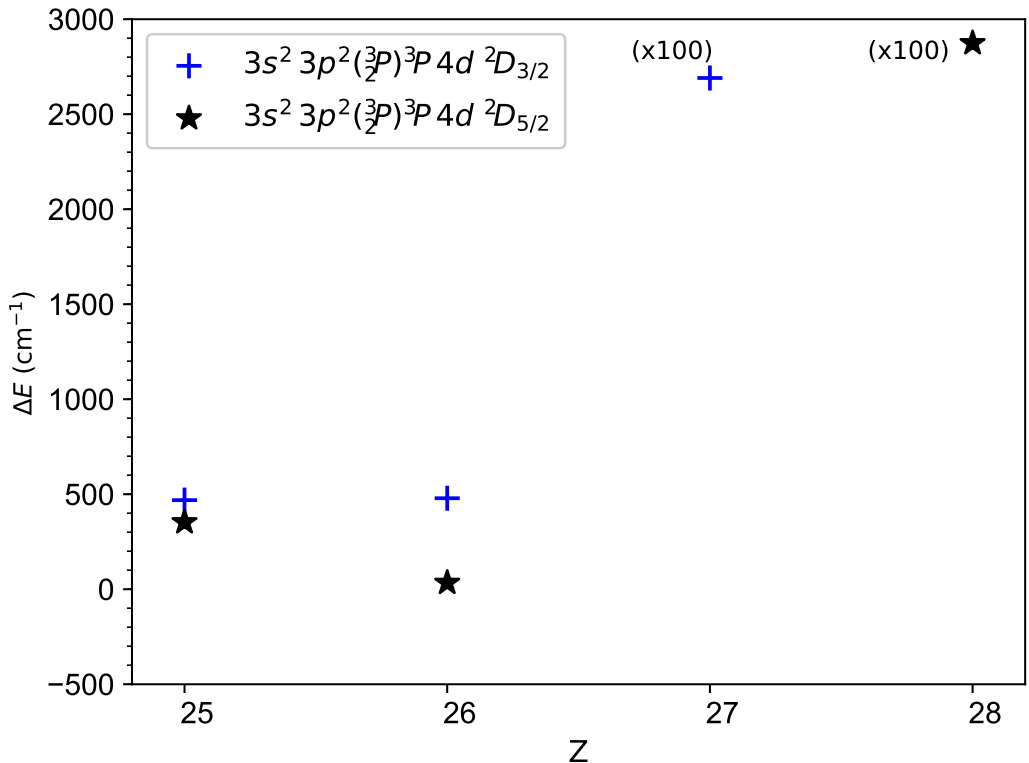}
	\caption{The energy differences $\Delta E$ (in cm$^{-1}$) between $E_{\rm MCDHF}$ and $E_{\rm NIST}$ are displayed as a function of the  nuclear charge $Z$  for the   $3s^{2}\,3p^{2}(^{3}_{2}P)~^{3}P\,4d~^{2}D_{3/2}$ and  $3s^{2}\,3p^{2}(^{3}_{2}P)~^{3}P\,4d~^{2}D_{5/2}$ states. The corresponding data are available in Table~\ref{tab.lev.all}. %The differences for the $3s^{2}\,3p^{2}(^{1}_{2}D)~^{1}D\,4s~^{2}D_{5/2}$ state in Mn~XI are reduced by 10 times from  its real values, which are labeled with ($\times 10$).  
		The real differences for the $3s^{2}\,3p^{2}(^{3}_{2}P)~^{3}P\,4d~^{2}D_{3/2}$ state in Co~XIII (Z=27) and  $3s^{2}\,3p^{2}(^{3}_{2}P)~^{3}P\,4d~^{2}D_{5/2}$  state in Ni~XIV (Z=28) are reduced by a factor of 100, as marked by the ($\times 100$) label.\label{figure.lev.nistlargedifferences}}
\end{figure*}

Our MCDHF excitation energies $E_{\rm MCDHF}$ (cm$^{-1}$), together with MCDHF radiative lifetimes $\tau_{\rm MCDHF}^l$ (in~s) in the length  form and $\tau_{\rm MCDHF}^v$ (in s)  in the velocity form, 
are reported in Table~\ref{tab.lev.all} for the lowest 546 (623, 701, 745) states of the $3p^4 3d$, $3s 3p^2 3d^2$, $3s 3p^3 4p$,  $3s 3p^4$, $3s^2 3d^3$, $3s^2 3p^2 3d$, $3s^2 3p^2 4d$, $3s^2 3p^2 4s$, $3p^3 3d^2$, $3p^5$, $3s 3p 3d^3$,  $3s 3p^3 3d$, $3s 3p^3 4s$, $3s^2 3p 3d^2$, %$3s^2 3p^2 4f$, 
$3s^2 3p^2 4p$, $3s^2 3p^3$ configurations in  Mn~XI (Fe~XII, Co~XIII, Ni~XIV).
Observed values $E_{\rm NIST}$   (cm$^{-1}$) from the NIST ASD \citep{Kramida.2018.V.p}, and energy differences $\Delta E$  (cm$^{-1}$) between   the values of $E_{\rm MCDHF}$ and $E_{\rm NIST}$ are also listed in this table. 
As many levels are strongly mixed, their configuration label is not unique.
  Here, the parity, $J$ value and energy  are used to match the levels to the NIST ASD.
  For those levels, the NIST values are shown in  blue color.
% GDZ: this is not necessary as it is a repetition.
% For the $n=3$ states, for which observed excitation energies are available, good agreement between $E_{\rm MCDHF}$ and $E_{\rm NIST}$  is obtained, with $\Delta E$ of only a few hundred cm$^{-1}$. 
The experimental excitation energies $E_{\rm NIST}$ agree well with our $E_{\rm MCDHF}$ values for a majority of the $n=4$ states.
However, there are 14 levels, including 4 levels in Mn~XI, 4 levels in Fe~XII, 4 levels in Co~XIII, and 2 levels in  Ni~XIV, for which differences are over 2~000 cm$^{-1}$. 
The $3s^{2}\,3p^{2}(^{3}_{2}P)~^{3}P\,4d~^{2}F_{7/2}$ state in Fe~XII,
where $E_{\rm NIST}$ and $E_{\rm MCDHF}$ differ by 3~616 cm$^{-1}$,  
is discussed below in detail. For the same levels, large difference of about  3~700 cm$^{-1}$
also occur for  Mn~XI  and Co~XIII.
As an example, the energy differences $\Delta E$ between $E_{\rm MCDHF}$ and $E_{\rm NIST}$
as a function of the  nuclear charge $Z$ are displayed in Figure~\ref{figure.lev.nistlargedifferences}
 for  the $3s^{2}\,3p^{2}(^{3}_{2}P)~^{3}P\,4d~^{2}D_{3/2}$, and  $3s^{2}\,3p^{2}(^{3}_{2}P)~^{3}P\,4d~^{2}D_{5/2}$ states. 
 Two anomalies appear for the $3s^{2}\,3p^{2}(^{3}_{2}P)~^{3}P\,4d~^{2}D_{3/2}$ state in Co~XIII
 and the  $3s^{2}\,3p^{2}(^{3}_{2}P)~^{3}P\,4d~^{2}D_{5/2}$  state in Ni~XIV. 
 Since  the same MCDHF computational  processes are carried out  for all the  ions, the accuracy of
 our MCDHF excitation energies  along the sequence should be systematic and consistency is expected. 
 Therefore, the large differences indicate that the identifications involving these levels
 are questionable or the wavelengths are incorrect.

\subsection{ Fe XII  Line Identifications }~\label{Sec:id}
In this section we review the identifications of the main Fe XII transitions 
$n=4 \to n=3$ transitions. 
Whilst \cite{DelZanna.2012.V543.p139} only considered
the strongest transitions at low plasma density (10$^{8}$ electrons cm$^{-3}$),
 here we consider the strongest transitions at high plasma density
 (10$^{19}$ electrons cm$^{-3}$), to review  Fawcett's identifications based on the
 laboratory plates. We have also considered one of Fawcett's plates for the 
evaluation.
Table~\ref{tab:lines} lists the strongest $n=4 \to n=3$ transitions in Fe XII,
calculated with the atomic data described in \cite{DelZanna.2012.V543.p139}. 
%{\it Note that the level description follows this latter calculation.}
%\mrg{\bf{(MRG: I don't understand the previous statement. What do you mean by ``level description''?)}}
In addition to Fawcett's wavelengths, we list the approximate {\sc autostructure}
values and our MCDHF wavelengths, with several comments.
Spectral lines that are strongest in low plasma density are noted with an asterisk.

%The transitions that are predicted to be strongest at high densities are from the 3s$^{2}$ 3p$^{2}$ 3d -- 3s$^{2}$ 3p$^{2}$ 4f array. Two of them should be visible at low densities. There are some discrepancies between the Fawcett's wavelengths and our predicted values, but the resolution of the laboratory spectra is insufficient to establish if some of the identifications are incorrect.

For the strongest lines of the 
$3s^{2} 3p^{2} 3d$ -- $3s^{2} 3p^{2} 4p$ transition array, we find a good
agreement (within 0.05~\AA) between Fawcett's wavelengths and our predicted values. 
%\mrg{({\bf Better than what? $\rightarrow$ I replaced ``better'' by ``good''})}
There are some important transitions visible at low densities, from the 
$\#392 / 3s^{2}\,3p^{2}(^{3}_{2}P)~^{3}P\,4p~^{4}S_{3/2}^{\circ}$ level. Our values are in excellent agreement (0.009~\AA) with the 
identifications proposed by \citet{DelZanna.2012.V546.p97}.
Generally, we confirm the  \mbox{CHIANTI1} experimental excitation energies
for the $3s^2 3p^2 4p$ levels ($\#371$, $\#380$, $\#387$,  $\#392$, $\#407$, $\#408$),
proposed by ~\citet{DelZanna.2012.V546.p97}, as 
the differences with our MCDHF1 are around a few hundreds cm$^{-1}$.
However, for  level $\#387 / 3s^{2}\,3p^{2}(^{3}_{2}P)~^{3}P\,4p~^{4}D_{7/2}^{\circ}$,
the difference  is larger ($-5~252$~cm$^{-1}$). The original assignment 
by \citet{Fawcett.1972.V5.p1255} of the line  at 108.44~\AA~value to the transition
$ 3s^{2}\,3p^{2}(^{3}_{2}P)~^{3}P\,4p~^{4}D_{7/2}^{\circ}$ --
$ 3s^{2}\,3p^{2}(^{3}_{2}P)~^{3}P\,3d~^{4}F_{9/2}$ is most likely correct. Using the
\mbox{CHIANTI1} experimental excitation energy 443~121 cm$^{-1}$ for the lower level, this results in
an energy of  1~365~290 cm$^{-1}$ for the upper level, in good agreement with our MCDHF1 result
(1~365~552 cm$^{-1}$). %\mrg{{(\bf PLEASE CHECK IF YOU ARE HAPPY WITH WHAT I WROTE!)}}

 Regarding the $3s^{2} 3p^{3}$ -- $3s^{2} 3p^{2} 4s$ transition array, very good 
 agreement with Fawcett's values is observed.
The differences of our MCDHF1 excitation energies and experimental values (NIST and \mbox{CHIANTI1})
for the $3s^2 3p^2 4s$ levels are within 300 cm$^{-1}$.

 We also confirm the 
identifications proposed by \citet{DelZanna.2012.V546.p97} of the decays from the \\$\#471 / 3s~^{2}S\,3p^{3}(^{4}_{3}S)~^{5}S\,4s~^{4}S_{3/2}^{\circ}$ level,
which are the strongest Fe XII transitions at low astrophysical densities. 
In \citet{DelZanna.2012.V546.p97}, the  83.336~\AA\ and 83.631~\AA\ lines observed by
  \citet{Behring.1972.V175.p493} were tentatively assigned to the decays
  from level  $\#471 / 3s~^{2}S\,3p^{3}(^{4}_{3}S)~^{5}S\,4s~^{4}S_{3/2}^{\circ}$
  to the lower levels $\#7 / 3s~^{2}S\,3p^{4}(^{3}_{2}P)~^{4}P_{3/2}$ and
  $\#8 / 3s~^{2}S\,3p^{4}(^{3}_{2}P)~^{4}P_{1/2}$, respectively.
  Our MCDHF1 wavelengths (83.335 \AA~and 83.636 \AA) for these two transitions show
  excellent agreement with the observations. 
  The difference between the MCDHF1 excitation energy (1~484~211 cm$^{-1}$) and the resulting
  \mbox{CHIANTI1} experimental energy (1~483~972 cm$^{-1}$) for the upper level is only $-$239 cm$^{-1}$.

  The situation for the experimental  energies of the $3s^2 3p^2 4d$ levels is  more complex
  and Fawcett's identifications of several  states are questionable,
  although we note that none of the transitions are strong in low-density astrophysical plasma.
As  pointed out by~\citet{DelZanna.2012.V546.p97}, the accuracy of previous theoretical wavelengths did not allow
firm identifications for the $3s^2 3p^2 4d$ levels.
We provide in   Table~\ref{tab:lines} suggestions for several revised identifications. 
The experimental excitation energies due to \citet{Fawcett.1972.V5.p1255}
for the $3s^2 3p^2 4d$ levels $\#498$, $\#507$, $\#552$,  $\#581$,
 have large deviations from our MCDHF1 results, with differences between
 $-$2~800 and  $-$5~200 cm$^{-1}$.
 
 As an example, we consider the  transition $3s^{2}\,3p^{2}(^{3}_{2}P)$ $^{3}P\,4d~^{2}F_{7/2}$ --
 $3s^{2}\,3p^{3}(^{2}_{3}D)~^{2}D_{5/2}^{\circ}$
 between levels $\#507$ and $\#3$. % (levels numbered 503 and 3 in Table~\ref{tab:lines}).
 This transition was assigned by 
 \cite{Fawcett.1972.V5.p1255} to the 67.702~\AA\ line, i.e. with 
 a wavelength  about 0.151~\AA\ greater than our MCDHF1 value (67.551~\AA). We have noted that
 the observed wavelength is very close to the MCDHF1 value (67.706~\AA) associated with the
  $ 3s^{2}\,3p^{2}(^{3}_{2}P)~^{3}P\,4d~^{2}F_{5/2}$ --  $3s^{2}\,3p^{3}(^{2}_{3}D)~^{2}D_{3/2}^{\circ}$ transition 
 between levels $\#498$ and $\#2$. %(numbered 491 and 2 in Table~\ref{tab:lines}). 
The transition rate for this latter ($\#498$--$\#2$) transition is $1.252\times 10^{11}$  s$^{-1}$,
about two times larger than the rate ($6.804\times 10^{10}$ s$^{-1}$) of the $\#507$--$\#3$ transition,
altough the two predicted intensities are very similar. 
Therefore, we suggest to assign the 67.702~\AA\ line to the transition
$\#498 / 3s^{2}\,3p^{2}(^{3}_{2}P)~^{3}P\,4d~^{2}F_{5/2}$ --  $\#2 / 3s^{2}\,3p^{3}(^{2}_{3}D)~^{2}D_{3/2}^{\circ}$.
As a consequence, the NIST  (and \mbox{CHIANTI1}) energy for the upper level $\#498$ should be changed to 1~518~627 cm$^{-1}$,
which agrees with our  MCDHF1 1~518~836   cm$^{-1}$ to within 210 cm$^{-1}$. 
Similar discrepancies are noted in Table~\ref{tab:lines}.

\subsection{Transition rates and lifetimes}

Wavelengths $\lambda_{ij}$ and the present MCDHF radiative transition data, which include transition
rates $A_{ji}$, weighted oscillator strengths $gf_{ji}$,  line strength $S_{ji}$, and branching fractions  (${\rm BF}_{ji} = A_{ji}/ \sum \limits_{k=1}^{j-1} A_{jk}$)  for electric-dipole (E1), magnetic dipole (M1), electric quadrupole (E2), and magnetic quadrupole (M2) transitions among all the levels listed in Table~\ref{tab.lev.all}  are reported in Table~ \ref{tab.trans.all}. 
E1 and E2 radiative transition data are given in both length ($l$) and velocity ($v$) forms.  
Using the uncertainty estimation approach~\citep{Kramida.2013.V63.p313,Kramida.2014.V212.p11}, for E1 and E2 transitions we provide  the estimated uncertainties of line strengths $S$ adopting the NIST ASD~\citep{Kramida.2018.V.p} terminology (A$^{+}$ $\leq$ 2~\%, A $\leq$ 3~\%, B$^{+}$ $\leq$ 7~\%, B $\leq$ 10~\%, C$^{+}$  $\leq$ 18~\%,  C $\leq$ 25~\%,  D$^{+}$ $\leq$ 40~\%, D $\leq$ 50~\%, and E $>$ 50~\% ) in the last column of this table.   
The difference $\delta S$ between line strengths $S_l$ and $S_v$ (in length and velocity forms respectively) is defined as $\delta S$ = $\left|S_{v}  - S_{l} \right|$/$\max (S_{v}$,~$S_{l})$. The averaged uncertainties $\delta S_{av}$ for line strengths $S$  for E1 transitions Fe~XII in various ranges of $S$ are assessed to 1~\% for $S \geq 10^{-1}$; 1.5~\% for $10^{-1} > S \geq 10^{-2}$; 2.7~\% for $10^{-2} > S \geq 10^{-3}$; 6~\% for $10^{-3} > S \geq 10^{-4}$; 12~\% for $10^{-4} > S \geq 10^{-5}$, and 23~\% for $10^{-5} > S \geq 10^{-6}$. Then, the larger of  $\delta S_{av}$ and $\delta S_{ji}$ is accepted as the uncertainty of each particular line strength. 
In Table~ \ref{tab.trans.all}, about 24~\% of E1 $S$ values in Fe~XII have uncertainties of  $\leq$ 2~\% (A+), 27~\% have uncertainties of  $\leq$ 3~\% (A), 29~\% have uncertainties of  $\leq$ 7~\% (B+), 2.4~\% have uncertainties of  $\leq$ 10~\% (B), 12~\% have uncertainties of  $\leq$ 18~\% (C+), 3.6~\% have uncertainties of  $\leq$ 25~\% (C), and 0.9~\% have uncertainties of  $\leq$ 40~\% (D+), while only 0.3~\% have uncertainties of  $>$ 40~\% (D and E).

In the spirit of the uncertainty estimation approach~\citep{Kramida.2013.V63.p313,Kramida.2014.V212.p11}, the estimated uncertainties of line strengths $S$ for E2  transitions in Fe~XII are estimated, as well as those for E1 and E2 transitions in Mn~XI, Co~XIII, and Ni~XIV. The estimated uncertainties for all E1 and E2 transitions with BF  $\geq 10^{-5}$ in Mn~XI, Fe~XII, Co~XIII, and Ni~XIV, are listed in Table~\ref{tab.trans.all}.

Our MCDHF radiative lifetimes $\tau_{\rm MCDHF}^l$ (in s) in the length  form and $\tau_{\rm MCDHF}^v$ (in s)  in the velocity form, for the lowest 546 (623, 701, 745) states of the $3p^4 3d$, $3s 3p^2 3d^2$, $3s 3p^3 4p$,  $3s 3p^4$, $3s^2 3d^3$, $3s^2 3p^2 3d$, $3s^2 3p^2 4d$, $3s^2 3p^2 4s$, $3p^3 3d^2$, $3p^5$, $3s 3p 3d^3$,  $3s 3p^3 3d$, $3s 3p^3 4s$, $3s^2 3p 3d^2$, %$3s^2 3p^2 4f$, 
$3s^2 3p^2 4p$, $3s^2 3p^3$ configurations in  Mn~XI (Fe~XII, Co~XIII, Ni~XIV), which are calculated by considering all possible E1, E2, M1, and M2 transitions, are provided in Table~\ref{tab.lev.all}. Our MCDHF radiative lifetimes $\tau_{\rm MCDHF}^l$ and  $\tau_{\rm MCDHF}^v$ show good agreement. For example, the average deviation between $\tau_{\rm MCDHF}^l$ and  $\tau_{\rm MCDHF}^v$ for all 623 levels in Fe~XII is 1~\%.

\subsection{Summary}
Using the MCDHF method combined with the RCI approach, including the transverse electron interaction in the low-frequency limit and the  leading QED effects  corrections,  calculations have been performed  for the lowest 546 (623, 701, 745) levels of the $3p^4 3d$, $3s 3p^2 3d^2$, $3s 3p^3 4p$,  $3s 3p^4$, $3s^2 3d^3$, $3s^2 3p^2 3d$, $3s^2 3p^2 4d$, $3s^2 3p^2 4s$, $3p^3 3d^2$, $3p^5$, $3s 3p 3d^3$,  $3s 3p^3 3d$, $3s 3p^3 4s$, $3s^2 3p 3d^2$, %$3s^2 3p^2 4f$, 
$3s^2 3p^2 4p$, $3s^2 3p^3$ configurations in  Mn~XI (Fe~XII, Co~XIII, Ni~XIV).
Excitation energies, radiative transition data, and  lifetimes  are reported. 

  Our detailed discussion of the excitation energies of the $n=4$ levels for  \ion{Fe}{12}, highlights that the identifications are questionable for a few  $n=4$ states.
  The comparison between experimental and predicted energies clearly shows that the present  calculations reach
  spectroscopic accuracy for these high-lying states.
  On that basis,  several identifications in the other isoelectronic ions are also uncertain. 
Our calculated excitation energies, as well as radiative transition data,  can be used to reliably identify the remaining Fe XII and Ni~XIV levels, and especially to identify all the $n=4$ states along the isoelectronic sequence of P-like ions,
where very little experimental data are available. 
The present MCDHF study should therefore stimulate further experimental investigations 
%with new level and spectral line identifications 
for those ions.  
The resulting accurate and consistent MCDHF data set will be useful for astrophysical modeling, line identification work and also for benchmarking other calculations.

\section*{Scientific software packages}
Scientific software packages including~\software{GRASP2K \citep{Jonsson.2007.V177.p597,Jonsson.2013.V184.p2197}
and	CHIANTI \citep{Dere.2019.V241.p22,Dere.1997.V125.p149}} are used in the present work. 	

\acknowledgments
We acknowledge the support from the National Key Research and Development Program of China under Grant No.~2017YFA0403200, the Science Challenge Project of China Academy of Engineering Physics (CAEP) under Grant No. TZ2016005, the National Natural Science Foundation of China (Grant No.~11703004, No.~11674066, No.~11504421, and No.~11734013), the Natural Science Foundation of Hebei Province, China (A2019201300 and A2017201165), and the Natural Science Foundation of Educational Department of Hebei Province, China (BJ2018058). This work is also supported by the Fonds de la Recherche Scientifique - (FNRS) and the Fonds Wetenschappelijk Onderzoek - Vlaanderen (FWO) under EOS Project n$^{\rm o}$~O022818F, and by the Swedish research council under contracts 2015-04842 and 2016-04185.  
GDZ acknowledges support form STFC (UK) via the consolidated grant to the solar/atomic astrophysics group, DAMTP, University of Cambridge.
KW expresses his gratitude to the support from the visiting researcher program at the Fudan University.

\clearpage
\bibliographystyle{aasjournal}
\bibliography{ref}
%\bibliography{../../../../../article/ref}

\clearpage
\startlongtable

\begin{deluxetable*}{clrrrrrrrrcc}
	\centering
	\tabletypesize{\tiny}
	\setlength{\tabcolsep}{3pt}
	\tablecaption{Excitation energies  $E$ in cm$^{-1}$ for 623 states of \ion{Fe}{12} from the present MCDHF calculations (hereafter referred to as MCDHF1).  
		For comparison, excitation energies from our recent MCDHF calculations~\citep{Wang.2018.V235.p27} (MCDHF2), experimental excitation energies compiled in the NIST ASD~\citep{Kramida.2018.V.p}, experimental (\mbox{CHIANTI1}) and calculated (CHIANTI2) values from the CHIANTI version 9~\citep{Dere.2019.V241.p22,Dere.1997.V125.p149} are also included, along with their differences ($\Delta E_x = E_x - E_{\rm MCDHF1}$) in cm$^{-1}$ with the present MCDHF values. 
		\label{tab.lev.fe}
	}
	\tablehead{\colhead{Key} & \colhead{Level} & \colhead{$E_{\rm MCDHF1}$} & \colhead{$E_{\rm MCDHF2}$}& \colhead{$\Delta E_{\rm MCDHF2}$}  & \colhead{$ E_{\rm NIST}$} & \colhead{$\Delta E_{\rm NIST}$} & \colhead{$E_{\rm{\mbox{CHIANTI1}}}$} &   \colhead{$\Delta E_{\rm{\mbox{CHIANTI1}}}$} &  \colhead{$E_{\rm{CHIANTI2}}$} & \colhead{$\Delta E_{\rm{CHIANTI2}}$}  }
	\startdata
1   & $3s^{2}\,3p^{3}(^{4}_{3}S)~^{4}S_{3/2}^{\circ}$                                              &0       &  0         &0     &    0      &   0      &   0              &               &          0       &  0            \\
2   & $3s^{2}\,3p^{3}(^{2}_{3}D)~^{2}D_{3/2}^{\circ}$                                              &41872   &  41771     &-101  &    41566  &   -306   &   41555.602      &     -316.398  &          41556   &  -316         \\
3   & $3s^{2}\,3p^{3}(^{2}_{3}D)~^{2}D_{5/2}^{\circ}$                                              &46386   &  46320     &-66   &    46075  &   -311   &   46088          &     -298      &          46088   &  -298         \\
4   & $3s^{2}\,3p^{3}(^{2}_{1}P)~^{2}P_{1/2}^{\circ}$                                              &74557   &  74373     &-185  &    74109  &   -448   &   74107          &     -450      &          74107   &  -450         \\
5   & $3s^{2}\,3p^{3}(^{2}_{1}P)~^{2}P_{3/2}^{\circ}$                                              &80926   &  80763     &-163  &    80515  &   -411   &   80515          &     -411      &          80515   &  -411         \\
6   & $3s~^{2}S\,3p^{4}(^{3}_{2}P)~^{4}P_{5/2}$                                                    &274618  &  274499    &-119  &    274373 &   -245   &   274373         &     -245      &          274373  &  -245         \\
7   & $3s~^{2}S\,3p^{4}(^{3}_{2}P)~^{4}P_{3/2}$                                                    &284240  &  284127    &-114  &    284005 &   -235   &   284005         &     -235      &          277374  &  -6866        \\
8   & $3s~^{2}S\,3p^{4}(^{3}_{2}P)~^{4}P_{1/2}$                                                    &288550  &  288452    &-98   &    288307 &   -243   &   288307         &     -243      &          288307  &  -243         \\
\enddata
\end{deluxetable*}

\clearpage

\startlongtable
\begin{deluxetable*}{crlrrrrrrl}
	\centering
\tabletypesize{\tiny}
\setlength{\tabcolsep}{3pt}
	\tablecaption{Excitation energies in cm$^{-1}$ %and radiative lifetimes (in s) 
		for the 546 (623, 701, 745) lowest levels of  Mn~XI (Fe~XII, Co~XIII, Ni~XIV), as well as \\ 
		radiative lifetimes (in s).
		$E_{\rm MCDHF}$ - our MCDHF excitation energies; $E_{\rm NIST}$  - observed values compiled in the NIST ASD \\ \citep{Kramida.2018.V.p}; $\Delta E$: energy differences (in cm$^{-1}$)  between   the values of $E_{\rm MCDHF}$ and $E_{\rm NIST}$.  
		$\tau_{\rm MCDHF}^l$ - our MCDHF lifetimes in \\length form;   $\tau_{\rm MCDHF}^v$ - our MCDHF lifetimes  in velocity form;  $LS$-composition -  the $LS$ eigenvector compositions. 
		Many levels are  strongly \\mixed, implying ambiguous configuration labels. In these cases, the parity, $J$, and energy, are used to match the theoretical levels with the \\
		atomic energy levels from  NIST ASD. For those levels, the NIST values are shown in the blue color. Energy differences  larger than 2~000 cm$^{-1}$ \\ are displayed in red color. 
\label{tab.lev.all}}
	\tablehead{\colhead{$Z$} & \colhead{Key} & \colhead{Level}  &\colhead{$J \pi$}  & \colhead{$E_{\rm MCDHF}$}  & \colhead{$E_{\rm NIST}$} & \colhead{$\Delta E$} & \colhead{$\tau_{\rm MCDHF}^l$} & \colhead{$\tau_{\rm MCDHF}^v$} &$LS$-composition }
	\startdata
25&  1     &$3s^{2}\,3p^{3}(^{4}_{3}S)~^{4}S_{3/2}^{\circ}$             & 3/2- &   0        & 0        & 0         &        &	        &      0.95 + 0.02~$3s~^{2}S\,3p^{3}(^{2}_{3}D)~^{3}D\,3d~^{4}S^{\circ}$                                                                                                        \\                                                                                                           
25&  2     &$3s^{2}\,3p^{3}(^{2}_{3}D)~^{2}D_{3/2}^{\circ}$             & 3/2- &   39688    & 39384    & 304       &3.65E$-$02&	3.65E$-$02&      0.87 + 0.09~$3s^{2}\,3p^{3}(^{2}_{1}P)~^{2}P^{\circ}$                                                                                                                    \\                                                                                                           
25&  3     &$3s^{2}\,3p^{3}(^{2}_{3}D)~^{2}D_{5/2}^{\circ}$             & 5/2- &   42986    & 42702    & 284       &7.49E$-$01&	7.49E$-$01&   0.96                                                                                                                                                                        \\                                                                                                           
25&  4     &$3s^{2}\,3p^{3}(^{2}_{1}P)~^{2}P_{1/2}^{\circ}$             & 1/2- &   69389    & 68945    & 444       &6.34E$-$03&	6.34E$-$03&   0.95                                                                                                                                                                        \\
\enddata
	\tablecomments{ Table~\ref{tab.lev.all} is published in its entirety in the machine-readable format. The results for the levels, for which the NIST values are available, are shown here for guidance regarding its form and content.}
\end{deluxetable*}

%\clearpage

\begin{deluxetable*}{llllllll}
	\tabletypesize{\tiny}
	\setlength{\tabcolsep}{4pt}
	\tablecaption{A list of the strongest lines from $n=4$ levels for Fe XII.  The relative intensities in column 3   $Int=N_{j} A_{ji}/N_{\rm e}$ are normalised (photons) and are calculated at high density (10$^{19}$ cm$^{-3}$), using the atomic data in	\cite{DelZanna.2012.V543.p139}.
		Column 4 gives the experimental wavelength $\lambda_{\rm exp}$ (\AA), as in  NIST (due to Fawcett) or 
		due to \cite{DelZanna.2012.V546.p97} ($^D$). We add a 'No' if we consider the 
		previous identification incorrect.  Column 5 gives the {\sc autostructure} (AS)
		wavelength $\lambda_{\rm AS}$, column 6 our present wavelength
		$\lambda_{\rm MCDHF}$ (with differences with the experimental ones in bracket) , while 
		the following column  ($\lambda_{\rm rev}$) reports the  new experimental wavelengths thet we propose.  
		The last column indicates if a line is strong in low-density astrophysical plasma (*), if it is 
		blended (sbl), if its identification is questionable (?) or new (N). \label{tab:lines} }
	\tablehead{\colhead{$i$--$j$}  & \colhead{Transition}  & \colhead{$Int$} &
	        \colhead{$\lambda_{\rm exp}$} & \colhead{$\lambda_{\rm AS}$} & \colhead{$\lambda_{\rm MCDHF}$}
		          & \colhead{$\lambda_{\rm rev}$} & Notes    }
	\startdata
18--387 & $3s^{2}\,3p^{2}(^{3}_{2}P)~^{3}P\,3d~^{4}F_{9/2}$                                                    --$3s^{2}\,3p^{2}(^{3}_{2}P)~^{3}P\,4p~^{4}D_{7/2}^{\circ}$                                            & 1.2                  &  108.440     & 107.04 &  108.38  (-0.06)  &          &     \\ 
25--408 & $3s^{2}\,3p^{2}(^{1}_{2}D)~^{1}D\,3d~^{2}G_{9/2}$                                                    --$3s^{2}\,3p^{2}(^{1}_{2}D)~^{1}D\,4p~^{2}F_{7/2}^{\circ}$                                            & 1.1                  &  110.591     & 109.20 &  110.527 (-0.06)  &          &     \\ 
16--380 & $3s^{2}\,3p^{2}(^{3}_{2}P)~^{3}P\,3d~^{4}F_{7/2}$                                                    --$3s^{2}\,3p^{2}(^{3}_{2}P)~^{3}P\,4p~^{4}D_{5/2}^{\circ}$                                            & 0.85                 &  108.605     & 107.16 &  108.552 (-0.05)  &          &     \\ 
24--407 & $3s^{2}\,3p^{2}(^{1}_{2}D)~^{1}D\,3d~^{2}G_{7/2}$                                                    --$3s^{2}\,3p^{2}(^{1}_{2}D)~^{1}D\,4p~^{2}F_{5/2}^{\circ}$                                            & 0.81                 &  110.732     & 109.34 &  110.676 (-0.06)  &          &     \\ 
19--386 & $3s^{2}\,3p^{2}(^{1}_{2}D)~^{1}D\,3d~^{2}F_{7/2}$                                                    --$3s^{2}\,3p^{2}(^{3}_{2}P)~^{3}P\,4p~^{4}P_{5/2}^{\circ}$                                            & 0.65                 &  -           & 107.53 &  108.963          &          &     \\ 
23--397 & $3s^{2}\,3p^{2}(^{3}_{2}P)~^{3}P\,3d~^{4}D_{7/2}$                                                    --$3s^{2}\,3p^{2}(^{3}_{2}P)~^{3}P\,4p~^{2}D_{5/2}^{\circ}$                                            & 0.63                 &  -           & 107.62 &  109.024          &          &     \\
\enddata
%	\tablecomments{ Table~\ref{tab.trans.all} is published in its entirety in the machine-readable format. Part of the values for Fe~XII are shown here for guidance regarding its form and content.}
\end{deluxetable*}

%\clearpage

\begin{deluxetable*}{lllllllllllll}
	\tabletypesize{\tiny}
	\setlength{\tabcolsep}{4pt}
	\tablecaption{Transition wavelengths $\lambda$ (in {\AA}), transition rates $A$ (in s$^{-1}$), weighted oscillator strengths $gf$, and line strengths $S$ (in a.u.) 
		between the states of  Mn~XI (Fe~XII, Co~XIII, Ni~XIV) listed in Table~\ref{tab.lev.all}. E1 and E2 transition data in both the length ($l$) and velocity ($v$) forms are provided. Type is the type of the multipole, BF is the branching fraction from the upper level. The last column (Acc.) represents the estimated accuracies of the $S$ values using the terminologies of the NIST ASD. Only transitions with BF  $\geq 10^{-5}$ are presented. \label{tab.trans.all} }
	\tablehead{\colhead{$Z$} & \colhead{$i$} & \colhead{$j$} &    
		\colhead{$\lambda$} & \colhead{Type}  &\colhead{BF} & \colhead{$A^l$} & \colhead{$gf^l$} & \colhead{$S^l$} & \colhead{$A^v$} & \colhead{$gf^v$} & \colhead{$S^v$}&  \colhead{Acc.}      }
	\startdata
26  &  1 &   2&  2.388218E$+$03 & M1&  9.992E$-$01&  5.417E$+$01&  1.853E$-$07  &1.094E$-$01 &           &          &           &      \\
26  &  1 &   2&  2.388218E$+$03 & E2&  8.153E$-$04&  4.420E$-$02&  1.512E$-$10  &1.227E$-$02 & 4.393E$-$02 & 1.502E$-$10&  1.219E$-$02&   B+ \\
26  &  1 &   3&  2.155836E$+$03 & M1&  6.796E$-$01&  2.073E$+$00&  8.667E$-$09  &4.621E$-$03 &           &          &           &      \\
26  &  1 &   3&  2.155836E$+$03 & E2&  3.473E$-$02&  1.060E$-$01&  4.430E$-$10  &2.644E$-$02 & 1.055E$-$01 & 4.408E$-$10&  2.631E$-$02&   B+ \\
26  &  1 &   4&  1.341248E$+$03 & M1&  7.257E$-$01&  1.889E$+$02&  1.019E$-$07  &3.380E$-$02 &           &          &           &      \\
26  &  1 &   4&  1.341248E$+$03 & E2&  3.788E$-$04&  9.862E$-$02&  5.319E$-$11  &7.644E$-$04 & 9.696E$-$02 & 5.230E$-$11&  7.515E$-$04&   B+ \\
26  &  1 &   5&  1.235704E$+$03 & M1&  5.511E$-$01&  3.449E$+$02&  3.159E$-$07  &9.652E$-$02 &           &          &           &      \\
26  &  1 &   6&  3.641427E$+$02 & E1&  9.769E$-$01&  1.593E$+$09&  1.900E$-$01  &2.278E$-$01 & 1.644E$+$09 & 1.961E$-$01&  2.351E$-$01&   B+ \\
\enddata
	\tablecomments{ Table~\ref{tab.trans.all} is published in its entirety in the machine-readable format. Part of the values for Fe~XII are shown here for guidance regarding its form and content.}
\end{deluxetable*}

\listofchanges
\end{document}